\documentclass[journal=acsodf,manuscript=article,layout=twocolumn]{achemso}
\usepackage[version=4]{mhchem}
\usepackage{graphicx}
\usepackage{xcolor}
\usepackage{multirow}
\usepackage{amsmath}
\usepackage{array}
\usepackage{stfloats}
\usepackage{placeins}

\title{First-Principles Investigation of Irida-Graphane: A Wide-Band-Gap Semiconductor Derived from Irida-Graphene}

\author{Djardiel S. Gomes}
\affiliation{Princesa Isabel Campus, Federal Institute of Education, Science and Technology of Para\'{i}ba (IFPB), Princesa Isabel, PB, 58755-000, Brazil.}
\alsoaffiliation{Materials Science Postgraduate Program, College UnB Planaltina, University of Bras\'{i}lia, Bras\'{i}lia, DF, 73380-900, Brazil.}

\author{Isaac M. F\'{e}lix}
\affiliation{Center for Agri-food Science and Technology, Federal University of Campina Grande, Pombal, PB, 58840-000, Brazil.}
\alsoaffiliation{Department of Physics, Federal University of Campina Grande, Campina Grande, PB, 58429-900, Brazil.}

\author{Marcelo L. Pereira Junior}
\email{marcelo.lopes@unb.br}
\affiliation{Department of Electrical Engineering, College of Technology, University of Bras\'{i}lia, Bras\'{i}lia, DF, 70910-900, Brazil.}
\alsoaffiliation{Materials Science Postgraduate Program, College UnB Planaltina, University of Bras\'{i}lia, Bras\'{i}lia, DF, 73380-900, Brazil.}

\begin{document}

	\begin{abstract}
		Opening a band gap in metallic two-dimensional carbon allotropes is an important step toward their use in electronic and optoelectronic devices, and full hydrogenation has proven to be an effective route to achieve it. Here, we propose irida-graphane, the fully hydrogenated derivative of the irida-graphene, a monolayer composed of three-, six-, and eight-membered rings, and investigate its structural, mechanical, electronic, and optical properties using density functional theory (DFT) at the PBE and HSE06 levels combined with phonon and \textit{ab initio} molecular dynamics (AIMD) simulations. All carbon atoms adopt \textit{sp}$^{3}$ hybridization, with the hydrogen atoms of the three- and six-membered rings bonded to opposite sides of a buckled carbon framework, and the lattice parameter contracts by 4.3\% with respect to irida-graphene. The hydrogenation energy of $-0.53$~eV/H indicates that the hydrogenation of irida-graphene from molecular hydrogen is exothermic. The absence of imaginary phonon frequencies confirms the dynamical stability of irida-graphane, and AIMD simulations show that its structure remains intact at temperatures up to 2000~K. Irida-graphane is mechanically isotropic, with an in-plane stiffness of approximately 156~N/m, a Poisson's ratio of approximately 0.1, and critical strains of approximately 20\%. Hydrogenation transforms the metallic parent structure into a semiconductor with an indirect band gap of 4.68~eV (PBE) and 6.0~eV (HSE06). Irida-graphane is transparent in the visible range, with reflectivity below 0.3\%, and absorbs strongly in the deep ultraviolet, with absorption coefficients of the order of 10$^{5}$--10$^{6}$~cm$^{-1}$. These results suggest irida-graphane as a candidate wide-gap material for ultraviolet optoelectronics and transparent coatings.
	\end{abstract}

	\section{Introduction}

	Two-dimensional (2D) materials have attracted considerable attention over the past two decades owing to their combination of structural, mechanical, electronic, and optical properties that differ substantially from those of their bulk counterparts \cite{Butler2013, Bhimanapati2015, Novoselov2016}. The isolation of graphene in 2004 \cite{PLACEHOLDER_graphene_discovery_1} marked a turning point in materials science, revealing a material with high carrier mobility, remarkable mechanical strength, and high thermal conductivity \cite{PLACEHOLDER_graphene_properties_1, PLACEHOLDER_graphene_properties_2, PLACEHOLDER_graphene_properties_3, PLACEHOLDER_graphene_Y2D_1}. However, the absence of a band gap in graphene limits its direct use in transistors and digital logic \cite{PLACEHOLDER_graphene_zero_gap_1}, which has motivated intensive efforts to design carbon-based 2D materials with tunable electronic properties.

	One of the most fruitful directions in this effort has been the exploration of 2D carbon allotropes with non-hexagonal ring topologies. Graphdiyne was synthesized in 2010 \cite{PLACEHOLDER_carbon_allotropes_2D_1}, and the subsequent years saw the theoretical prediction of penta-graphene \cite{PLACEHOLDER_carbon_allotropes_2D_2}, phagraphene \cite{PLACEHOLDER_carbon_allotropes_2D_3}, and $\psi$-graphene \cite{PLACEHOLDER_carbon_allotropes_2D_7}, among many others. The synthesis of the biphenylene network in 2021 \cite{PLACEHOLDER_carbon_allotropes_2D_4} and of the monolayer fullerene network in 2022 \cite{PLACEHOLDER_carbon_allotropes_2D_5} demonstrated that 2D carbon networks beyond graphene are experimentally accessible, and renewed the interest in the theoretical design of new members of this family.

	In this context, irida-graphene, a planar all-\textit{sp}$^{2}$ carbon allotrope composed of fused three-, six-, and eight-membered rings \cite{Lu2013, PLACEHOLDER_irida_graphene_original_1}, was fully characterized in 2023, crystallizing in the hexagonal $P6/mmm$ space group with a lattice parameter of 6.34~\AA\ and a formation energy of $-6.96$~eV/atom \cite{PLACEHOLDER_irida_graphene_original_1}. Irida-graphene was shown to be dynamically stable and to preserve its planar structure in AIMD simulations up to 1000~K, and it is a nonmagnetic metal with a Dirac cone above the Fermi level and an elastic modulus of approximately 396~GPa \cite{PLACEHOLDER_irida_graphene_original_1}. Since then, irida-graphene has attracted considerable attention. Its porous network has been explored for hydrogen storage upon decoration with transition, alkali, and alkaline-earth metals \cite{Tan2024, Bhat2025, Yuan2024, PLACEHOLDER_irida_graphene_followups_3, Sun2025, Duan2024, ZhangY2024}, and for electrochemical energy storage as an anode for metal-ion batteries \cite{PLACEHOLDER_irida_graphene_followups_1, PLACEHOLDER_irida_graphene_followups_2, Kaur2024, Wei2024, Rusho2025}, an anchoring material for lithium-sulfur batteries \cite{Martins2025}, and an electrode for supercapacitors \cite{Rodrigues2026}. Its fracture behavior \cite{PLACEHOLDER_irida_graphene_mechanics_1} and phonon thermal transport \cite{PLACEHOLDER_irida_graphene_followups_5} have also been investigated, and its topology has inspired related carbon structures \cite{Majidi2024a, Majidi2024b, PLACEHOLDER_irida_graphene_followups_7} as well as boron nitride \cite{PLACEHOLDER_irida_graphene_followups_4} and silicon \cite{PLACEHOLDER_irida_graphene_followups_6} analogues. Notably, the metallic character of irida-graphene, which favors its use in electrochemical energy storage, also restricts its applicability in devices that require a band gap.

	Chemical functionalization provides a well-established route to open a band gap in 2D carbon materials \cite{PLACEHOLDER_sp3_functionalization_tuning_2}. The prototypical example is graphane, the fully hydrogenated form of graphene, which was predicted in 2007 \cite{PLACEHOLDER_graphane_prediction_1}, and experimental evidence for its formation through the reversible hydrogenation of graphene was reported in 2009 \cite{PLACEHOLDER_graphane_experiment_1}. In graphane, the conversion of all carbon atoms to \textit{sp}$^{3}$ hybridization removes the $\pi$ states and opens a direct band gap of approximately 3.5~eV at the PBE level \cite{PLACEHOLDER_graphane_prediction_1, PLACEHOLDER_graphane_gap_1}, which increases to 5.4~eV within the GW approximation \cite{PLACEHOLDER_graphane_gap_1}. Fluorination of graphene similarly yields a wide-gap insulator \cite{PLACEHOLDER_sp3_functionalization_tuning_1}. This strategy has been extended to other 2D carbon allotropes. The hydrogenation of penta-graphene widens its band gap \cite{PLACEHOLDER_hydrogenation_other_allotropes_1}, while the hydrogenation of the metallic biphenylene network \cite{PLACEHOLDER_hydrogenation_other_allotropes_2} and $\psi$-graphene \cite{PLACEHOLDER_hydrogenation_other_allotropes_4} opens band gaps whose magnitude depends on the hydrogen coverage \cite{PLACEHOLDER_hydrogenation_other_allotropes_4}.

	Despite the growing body of work on irida-graphene, its hydrogenated counterpart remains unexplored. To the best of our knowledge, no previous study has addressed whether the full hydrogenation of irida-graphene yields a stable monolayer, nor how its three-membered rings, which are highly strained in the \textit{sp}$^{2}$ lattice, accommodate the conversion to \textit{sp}$^{3}$ hybridization. These questions are relevant because the strain stored in the parent lattice may favor hydrogenation energetically, while the coexistence of rings of different sizes may lead to a buckling pattern and to physical properties distinct from those of graphane.

	In this work, we investigate the structural, mechanical, electronic, and optical properties of irida-graphane, the fully hydrogenated derivative of irida-graphene, using density functional theory (DFT) calculations at the PBE and HSE06 levels, complemented by phonon calculations and \textit{ab initio} molecular dynamics (AIMD) simulations. We show that irida-graphane is dynamically stable, remains structurally intact at temperatures up to 2000~K, energetically favored with respect to irida-graphene and molecular hydrogen, mechanically isotropic, and a wide-gap semiconductor that is transparent in the visible range and absorbs strongly in the deep ultraviolet.

	\section{Methodology}

	First-principles calculations were performed using the SIESTA code \cite{Soler2002, Garcia2020}, which is based on the DFT formalism \cite{Kohn1965, Sanchez1997}. The exchange-correlation effects were described using the generalized gradient approximation (GGA) with the Perdew-Burke-Ernzerhof (PBE) functional \cite{Perdew1996}. Given that GGA/PBE systematically underestimates the electronic band gap, additional calculations of the electronic and optical properties were carried out using the hybrid Heyd-Scuseria-Ernzerhof (HSE06) functional \cite{Heyd2003} as implemented in the HONPAS package \cite{Qin2015, Shang2020}. The electron-ion interactions were described by Troullier-Martins norm-conserving pseudopotentials \cite{Troullier1991} in the Kleinman-Bylander factored form \cite{Kleinman1982}, with valence electron configurations of 2\textit{s}$^{2}$2\textit{p}$^{2}$ and 1\textit{s}$^{1}$ for C and H atoms, respectively. A mesh cutoff of 800~Ry and a double-$\zeta$ polarized (DZP) basis set composed of numerical atomic orbitals with finite range were employed in all calculations. The Brillouin zone of the unit cell was sampled using a $40 \times 40 \times 1$ Monkhorst-Pack $k$-point mesh \cite{Monkhorst1976}. In the structural optimization, both the lattice vectors and atomic positions were fully relaxed until the maximum force acting on each atom was less than $10^{-3}$~eV/\AA\ and the difference in total energy was less than $10^{-5}$~eV. A simulation cell length of 30~\AA\ was adopted along the direction perpendicular to the monolayer, which corresponds to a vacuum region of approximately 27~\AA\ and prevents spurious interactions between periodic images.

	Phonon dispersion calculations were performed to assess the dynamical stability of the irida-graphane monolayer. A $3 \times 3 \times 1$ supercell was employed with an energy convergence criterion of $10^{-5}$~eV, a force convergence criterion of 0.001~eV/\AA, and a mesh cutoff of 800~Ry. The acoustic sum rule was applied to the vibrational frequencies at the $\Gamma$ point. Additionally, AIMD simulations were performed using a $2 \times 2 \times 1$ supercell containing 96 atoms to assess the thermal stability of irida-graphane at 300, 1000, and 2000~K. The simulations were performed within the NVT ensemble with a $10 \times 10 \times 1$ $k$-point mesh, using a time step of 1.0~fs and a Nos\'{e}-Hoover thermostat for temperature regulation \cite{PLACEHOLDER_nose_hoover_1, PLACEHOLDER_nose_hoover_3}, and the system was evolved for 10~ps at each temperature \cite{tromer2020, montes2016, silva2021}.

	The mechanical response was evaluated by applying uniaxial and biaxial tensile strains to a $2 \times 2 \times 1$ supercell in increments of 0.5\% up to structural failure. Uniaxial strain was applied independently along the $x$- and $y$-directions, while the supercell was allowed to relax along the direction perpendicular to the loading. The effective thickness of the monolayer, required to express the two-dimensional mechanical quantities in three-dimensional units, was obtained from the interlayer distance of stacked irida-graphane layers, and the calculations for bilayer, trilayer, and periodically stacked systems consistently yielded 5.50~\AA. Finally, the optical properties were computed for unpolarized light incident along the direction normal to the monolayer, so that the electric field lies in the basal plane and the response corresponds to an average over the in-plane directions, using a $10 \times 10 \times 1$ $k$-point mesh and a Gaussian broadening of 0.05~eV.

	\section{Results and Discussion}

\begin{figure}[h!]
		\centering
		\includegraphics[width=\linewidth]{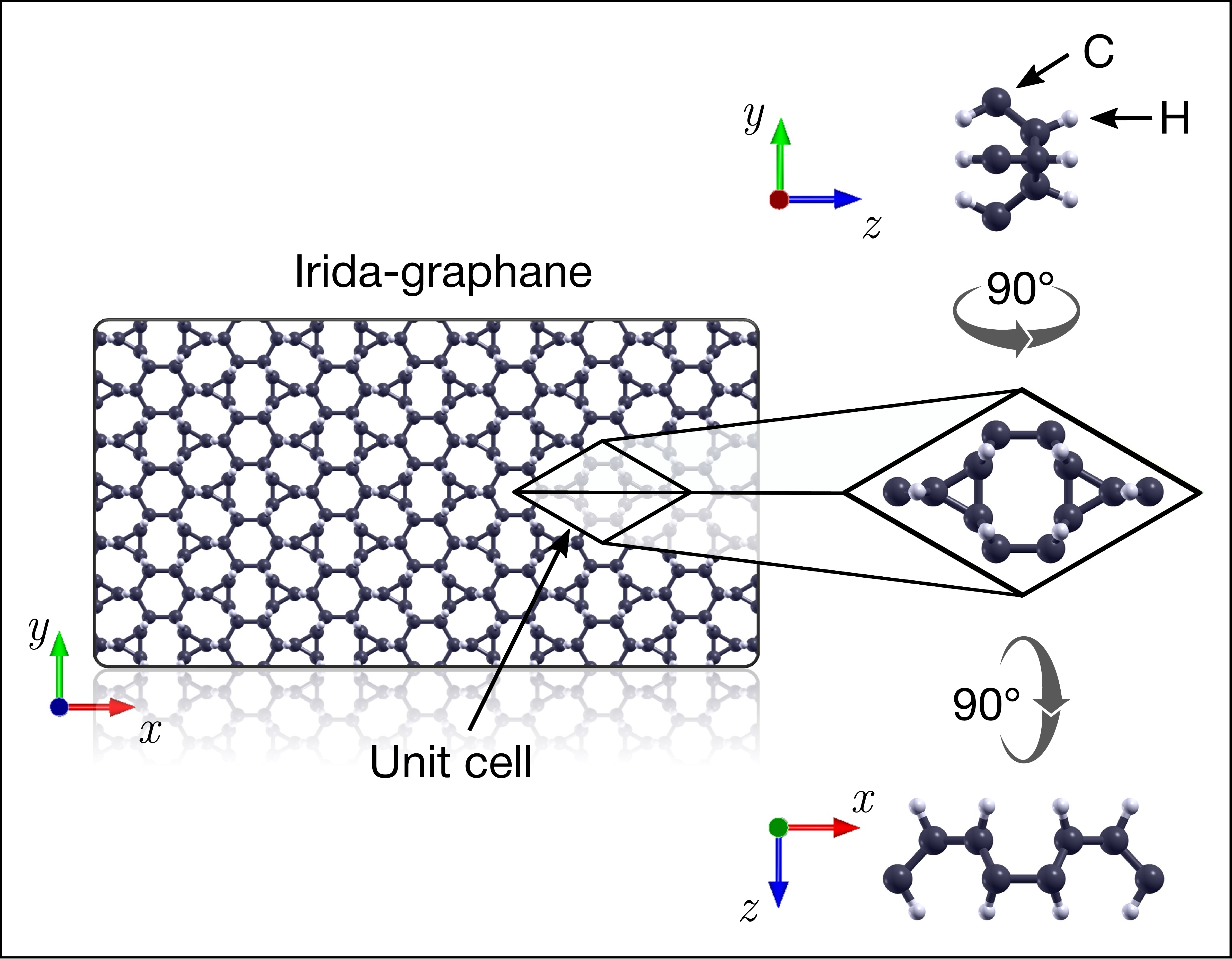}
		\caption{Atomic structure of the irida-graphane monolayer. The main panel shows the top view of an extended sheet, the inset shows the unit cell, and the side panels show views along the $x$- and $y$-directions. Gray and white spheres represent C and H atoms, respectively.}
		\label{fig:struct}
	\end{figure}

	Full hydrogenation transforms all carbon atoms of the parent lattice from \textit{sp}$^{2}$ to \textit{sp}$^{3}$ hybridization, and the way in which the three-, six-, and eight-membered rings accommodate this conversion determines the geometry of the resulting monolayer. The optimized structure of irida-graphane is shown in Figure~\ref{fig:struct}. Its hexagonal unit cell contains 24 atoms (12 C and 12 H) and has an optimized lattice parameter of 6.081~\AA, which is 4.3\% smaller than the value of 6.357~\AA\ obtained for pristine irida-graphene calculated using the same computational setup, in close agreement with the value of 6.34~\AA\ reported previously \cite{PLACEHOLDER_irida_graphene_original_1}. Each carbon atom is bonded to three carbon atoms and one hydrogen atom, confirming the complete conversion to \textit{sp}$^{3}$ hybridization in the lattice. The hydrogen atoms bonded to the carbon atoms of the three-membered rings point toward one side of the monolayer, while those bonded to the carbon atoms of the six-membered rings point toward the opposite side. This arrangement induces, the carbon framework becomes buckled, resulting in a vertical separation of 1.122~\AA\ between the two carbon sublayers and a total thickness of 3.077~\AA\ measured between the outermost hydrogen atoms.

    This arrangement is reflected in the C–C bond lengths. Within the three-membered rings, the C–C bonds connect carbon atoms whose hydrogen atoms lie on the same side of the monolayer and measure 1.521~\AA. The bonds connecting the three- and six-membered rings, where the bonded carbon atoms bear hydrogen atoms on opposite sides of the sheet, are slightly longer, measuring 1.529–1.530~\AA. Within the six-membered rings, the C–C bonds reach 1.593~\AA, with the hydrogen atoms again lying on the same side of the monolayer. Thus, along the carbon framework, the hydrogen atoms alternate in pairs rather than individually. This pattern combines chair-like arrangements between the rings with boat-like arrangements within them, in contrast to the uniform chair conformation of graphane \cite{PLACEHOLDER_graphane_prediction_1}.

    All these C–C bonds are longer than those calculated for pristine irida-graphene, for which we obtain 1.417~\AA\ within the three-membered rings and 1.408–1.444~\AA\ for the remaining bonds. The resulting bond lengths are broadly consistent with single C–C bonds, whose characteristic length is approximately 1.54~\AA\ in diamond \cite{PLACEHOLDER_diamond_cc_1}. Notably, the C–C bond length within the three-membered rings is close to the equilibrium value of 1.503~\AA\ reported for cyclopropane \cite{PLACEHOLDER_cyclopropane_cc_1}, suggesting that the triangular motifs of the parent lattice are retained as saturated, cyclopropane-like units. In contrast, the longer C–C bonds within the six-membered rings may be associated with steric repulsion between hydrogen atoms attached to adjacent carbon atoms on the same side of the monolayer. These hydrogen atoms may experience enhanced repulsion because of their eclipsed arrangement, imposed by the ring geometry.

    The C–H bond lengths range from 1.103 to 1.110~\AA. The bond angles around the carbon atoms span 104.6–126.1$^{\circ}$, excluding the 60$^{\circ}$ angles within the three-membered rings. This broad angular range indicates a substantial distortion of the local tetrahedral geometry associated with the \textit{sp}$^{3}$-hybridized carbon atoms.
   
	Interestingly, the lattice parameter contracts upon hydrogenation despite the elongation of all C–C bonds: by 7.3\% within the three-membered rings, 8.6\% between the three- and six-membered rings, and 10.3\% within the six-membered rings. This apparent contradiction can be explained by the buckling of the carbon framework, which tilts the bonds out of the basal plane and thereby reduces their in-plane projections. This behavior contrasts with that of graphene, whose calculated lattice parameter increases from 2.46~\AA\ to approximately 2.52–2.54~\AA\ upon full hydrogenation to chair graphane \cite{PLACEHOLDER_graphane_prediction_1, PLACEHOLDER_graphane_hydrogenation_energy_1}.

    The energetic favorability of hydrogenation provides a further criterion for assessing the feasibility of obtaining irida-graphane from irida-graphene. To quantify this contribution, we calculated the hydrogenation energy per hydrogen atom relative to molecular hydrogen, defined as
    \begin{equation}
    E_{\text{H}} =
    \frac{E_{\text{IGH}}-E_{\text{IG}}-\frac{n_{\text{H}}}{2}E_{\text{H}_2}}
    {n_{\text{H}}} ,
    \end{equation}
    \noindent where $E_{\text{IGH}}$ and $E_{\text{IG}}$ are the total energies of the irida-graphane and irida-graphene unit cells, respectively, $E_{\text{H}_2}$ is the total energy of an isolated H$_2$ molecule, and $n_{\text{H}}=12$ is the number of hydrogen atoms in the unit cell. We obtain $E_{\text{H}}=-0.53$~eV/H, indicating that hydrogenation is exothermic with respect to irida-graphene and molecular hydrogen at the level of theory employed. This value is more negative than the corresponding value of approximately $-0.19$~eV/H inferred from the PBE formation energy of $-0.097$~eV/atom reported for chair graphane relative to graphene and H$_2$ \cite{PLACEHOLDER_graphane_hydrogenation_energy_1}. The greater exothermicity found for irida-graphane may be associated with the release of strain in the \textit{sp}$^{2}$-bonded three-membered rings of the parent lattice as they transform into saturated, cyclopropane-like units.

\begin{figure}[h!]
		\centering
		\includegraphics[width=0.95\linewidth]{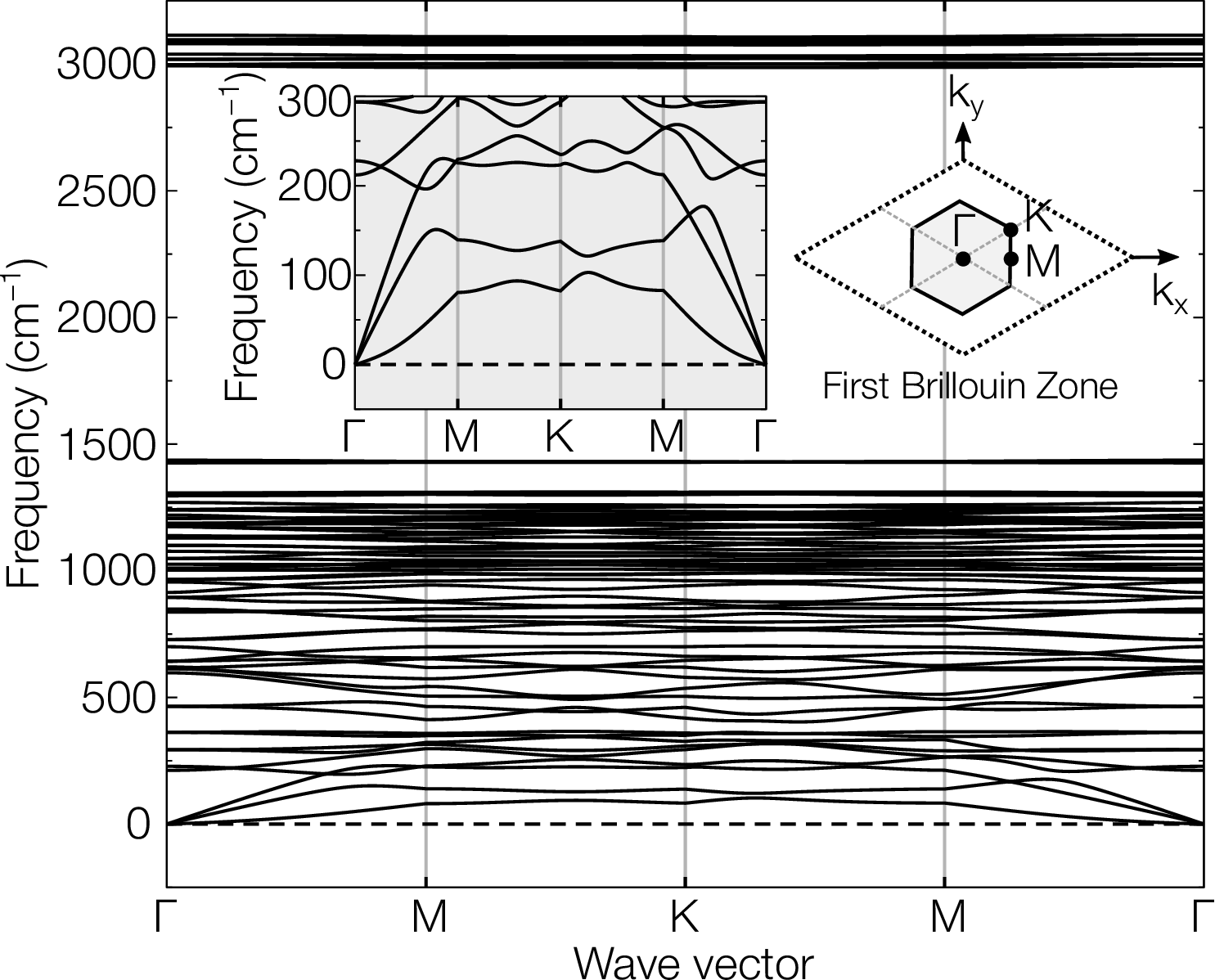}
		\caption{Phonon dispersion of the irida-graphane monolayer along the $\Gamma$-M-K-M-$\Gamma$ path. The inset shows the low-frequency region and the first Brillouin zone.}
		\label{fig:dispersion}
	\end{figure}

	A favorable hydrogenation energy, however, does not ensure that the resulting lattice is stable against small atomic displacements. To assess its dynamical stability, we have calculated the phonon dispersion of irida-graphane along the $\Gamma$-M-K-M-$\Gamma$ path of the first Brillouin zone, as presented in Figure~\ref{fig:dispersion}. The 24 atoms of the unit cell give rise to 72 phonon branches, of which three are acoustic and 69 are optical. We observe no imaginary frequencies along the sampled path, supporting the dynamical stability of the monolayer. As shown in the inset, the longitudinal and transverse acoustic branches exhibit linear dispersion from the $\Gamma$ point, while the out-of-plane acoustic branch exhibits the quadratic dispersion characteristic of the flexural mode of two-dimensional materials. No pronounced softening is observed near the zone center, and the lowest optical mode at $\Gamma$ lies at 212~cm$^{-1}$.

	At higher frequencies, the vibrational spectrum splits into well-separated groups of modes, with three main frequency gaps at 366--408~cm$^{-1}$, 1311--1426~cm$^{-1}$, and 1436--2987~cm$^{-1}$. The widest one, of approximately 1550~cm$^{-1}$, isolates the C-H stretching modes, which form a narrow group of bands between 2987 and 3115~cm$^{-1}$. These modes are nearly dispersionless, as expected for localized vibrations of the light hydrogen atoms against the much heavier carbon framework. A similarly wide gap separating the C-H stretching modes from the framework modes is present in the phonon dispersion of graphane, whose C-H stretching frequencies at the $\Gamma$ point were calculated at 2739 and 2783~cm$^{-1}$ \cite{PLACEHOLDER_graphane_phonons_1}. Below approximately 1440~cm$^{-1}$, the modes involve the vibrations of the carbon framework coupled to the C-H bending motions.

\begin{figure*}[h!]
		\centering
		\includegraphics[width=0.85\linewidth]{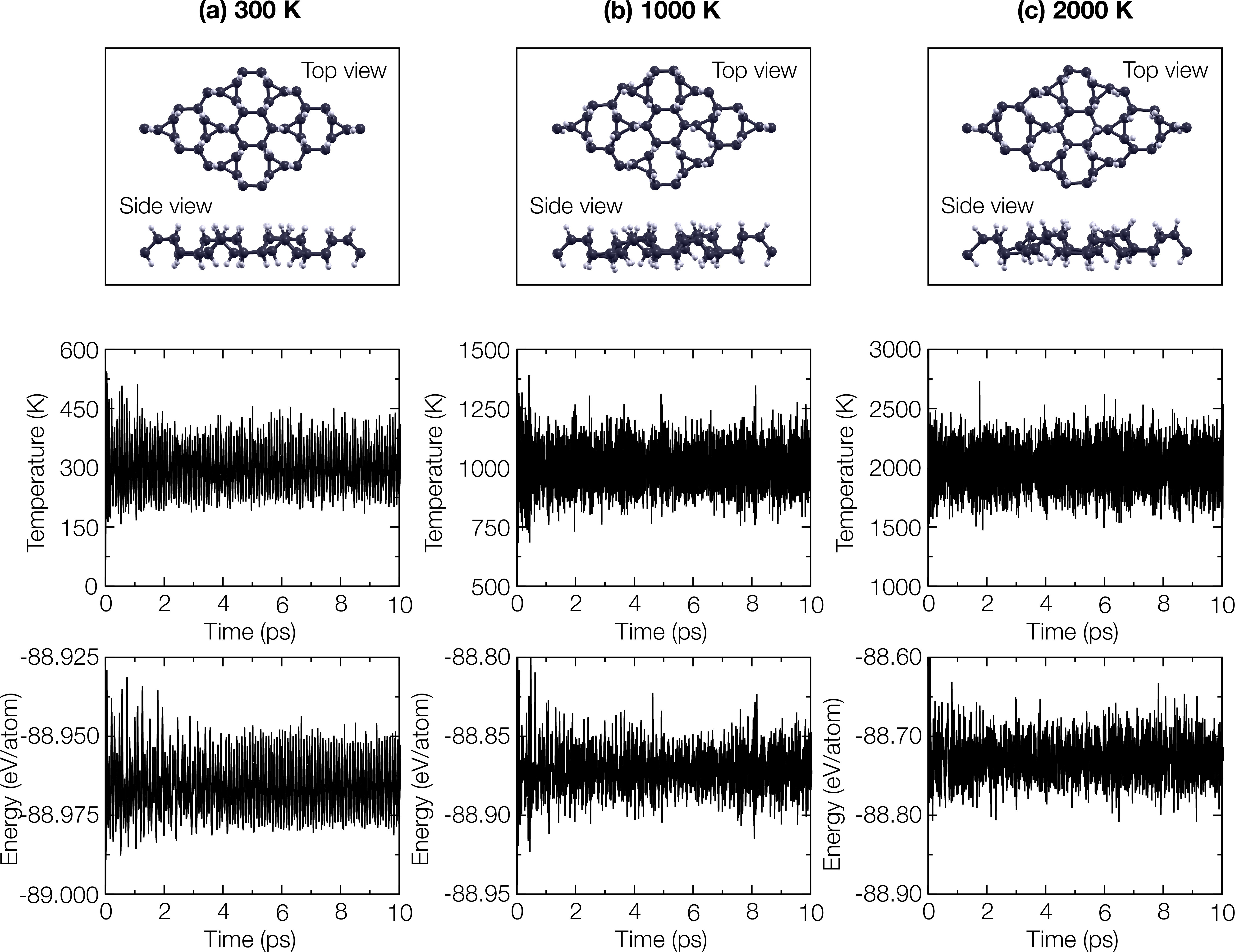}
		\caption{AIMD simulations of irida-graphane at (a) 300~K, (b) 1000~K, and (c) 2000~K. The top row shows the final snapshots (top and side views), and the middle and bottom rows show the time evolution of the temperature and of the total energy per atom, respectively.}
		\label{fig:aimd}
	\end{figure*}

	Dynamical stability at zero temperature, in turn, does not guarantee that the lattice survives thermal fluctuations, particularly in a hydrogenated structure in which hydrogen desorption or ring reconstruction could be thermally activated. For this reason, the thermal stability of irida-graphane was evaluated through AIMD simulations at 300, 1000, and 2000~K, as presented in Figure~\ref{fig:aimd}. At 300~K, the temperature fluctuates around an average value of $300.0 \pm 60.4$~K, and the Kohn-Sham total energy fluctuates around $-88.966 \pm 0.009$~eV/atom with no systematic drift over the 10~ps trajectory. A similar behavior is observed at 1000 and 2000~K, in which the energy oscillates around constant mean values after the initial thermalization, with larger fluctuation amplitudes that follow the increase in temperature. The snapshots at the end of the simulations show that the carbon framework retains its three-, six-, and eight-membered rings at all temperatures, with thermal fluctuations of the atomic positions that increase with temperature and no evidence of bond breaking, ring reconstruction, or hydrogen desorption. Together with the phonon results, these simulations indicate that irida-graphane is dynamically stable and thermally robust up to at least 2000~K within the simulated time scale.

\begin{figure*}[h!]
		\centering
		\includegraphics[width=0.75\linewidth]{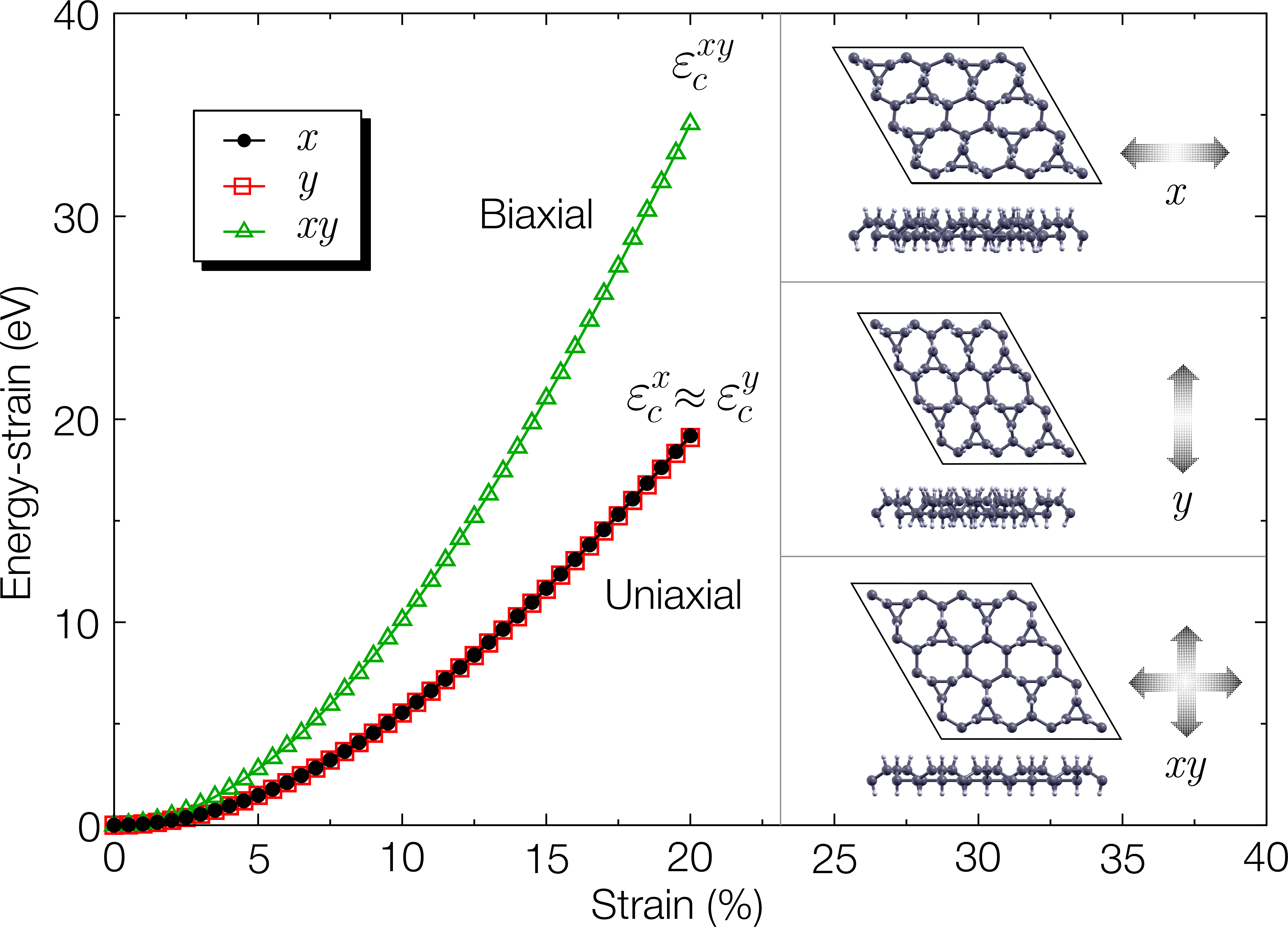}
		\caption{Strain energy of the irida-graphane supercell as a function of uniaxial strain along the $x$- and $y$-directions and of biaxial strain ($xy$). The critical strains are indicated, and the right panels show the structures under the corresponding strains.}
		\label{fig:strain}
	\end{figure*}

	Once its stability is established, a natural question is how the conversion of the \textit{sp}$^{2}$ network into a buckled \textit{sp}$^{3}$ network with longer bonds affects the response of the lattice to in-plane loading. We have addressed it by applying uniaxial strain along the $x$- and $y$-directions and biaxial strain along both directions simultaneously, and the resulting strain energy curves are presented in Figure~\ref{fig:strain}. The curves for uniaxial strain along the $x$- and $y$-directions are practically superimposed over the entire strain range, which reflects the in-plane mechanical isotropy expected for a lattice with hexagonal symmetry. Since the strain energy is stored in both directions simultaneously, the biaxial curve rises more steeply and reaches 34.6~eV per supercell at 20\% strain, compared with approximately 19.2~eV for uniaxial strain.

	From the quadratic region of these curves, we have extracted the in-plane stiffness \cite{Ozcelik2013, Topsakal2010}
	\begin{equation}
		C=\frac{1}{A} \times\left.\left(\frac{\partial^2 E_S}{\partial \varepsilon^2}\right)\right|_{\varepsilon\;=\;0} ,
	\end{equation}
	\noindent where $A$ is the area of the optimized supercell, $\varepsilon = \Delta l/l_0$ is the uniaxial strain with $l_0$ being the equilibrium supercell length and $\Delta l = l - l_0$, and $E_S$ is the strain energy obtained by subtracting the equilibrium total energy from the total energy under strain, as well as the in-plane bulk modulus \cite{Majidi2017, Asadpour2015}
	\begin{equation}
		B=A \times\left.\left(\frac{\partial^2 E_S}{\partial A^2}\right)\right|_{A_{0}} ,
	\end{equation}
	\noindent where $A_0$ and $A$ represent the supercell area at equilibrium and under biaxial strain, respectively. We obtain an in-plane stiffness of approximately 156~N/m along both the $x$- and $y$-directions and an in-plane bulk modulus of 77~N/m. Using the effective thickness of 5.50~\AA, the in-plane stiffness corresponds to a Young's modulus of approximately 284~GPa. These values are considerably lower than the in-plane stiffness of 340~N/m for graphene \cite{PLACEHOLDER_graphene_Y2D_1} and of 243--246~N/m for graphane \cite{Topsakal2010, PLACEHOLDER_graphane_mechanics_2}, which is consistent with the lower areal density of bonds in a lattice containing eight-membered rings and with the replacement of \textit{sp}$^{2}$ bonds by longer \textit{sp}$^{3}$ single bonds. In contrast, the elastic modulus of approximately 396~GPa reported for irida-graphene from reactive molecular dynamics simulations \cite{PLACEHOLDER_irida_graphene_original_1} corresponds to approximately 133~N/m for a thickness of 3.35~\AA, which indicates that, despite the longer bonds, the full hydrogenation does not reduce the in-plane stiffness of the parent lattice and may even increase it slightly. We note that this comparison is made in terms of two-dimensional quantities, since the values in GPa depend on the thickness adopted for each monolayer. This behavior may be attributed to the additional rigidity introduced by the buckled \textit{sp}$^{3}$ framework, although a quantitative comparison is limited by the different levels of theory employed in the two studies. The lateral response was quantified through the Poisson's ratio \cite{Kang2011}
	\begin{equation}
		\nu=-\frac{\varepsilon_{\text{trans}}}{\varepsilon_{\text{axial}}} ,
	\end{equation}
	\noindent obtained from the ratio of the transverse strain ($\varepsilon_{\text{trans}}$) to the axial strain ($\varepsilon_{\text{axial}}$) at small strains, which yields $\nu_{x}$ = 0.08 and $\nu_{y}$ = 0.11 for strain applied along the $x$- and $y$-directions, respectively. Such small values are comparable to the values of 0.07--0.08 reported for graphane \cite{Topsakal2010, PLACEHOLDER_graphane_mechanics_2} and indicate that the transverse contraction under uniaxial tension is weak, which may be attributed to the buckled \textit{sp}$^{3}$ framework, in which the applied strain is accommodated mainly by the flattening of the buckled carbon sublayers and the opening of the bond angles rather than by a lateral contraction of the lattice.

	Beyond the elastic regime, the in-plane stress, obtained from the stress tensor and converted into a two-dimensional quantity by multiplying it by the length of the simulation cell normal to the monolayer, increases monotonically with strain up to the critical strain, which we identify as the strain at which the stress reaches its maximum value. The critical strains are approximately 20\% for uniaxial strain along both directions and for biaxial strain, as indicated in Figure~\ref{fig:strain}, and the corresponding ultimate strengths are 17.5~N/m along the $x$-direction, 16.7~N/m along the $y$-direction, and 15.1~N/m for biaxial strain, which correspond to 31.8, 30.4, and 27.5~GPa, respectively. Under biaxial strain, the stress collapses immediately beyond the critical strain, at 21\%, signaling the simultaneous rupture of C-C bonds and the structural failure of the monolayer. Under uniaxial strain, in contrast, the stress decreases slowly up to 21.5\%, drops by approximately 20\% between 21.5 and 22\%, declines gradually to approximately 6--7~N/m at 24.5\%, and only then collapses at 25\%, which suggests that the lattice accommodates the deformation through successive local bond ruptures before complete failure. Although these ultimate strengths are well below the intrinsic strength of 42~N/m measured for graphene \cite{PLACEHOLDER_graphene_Y2D_1}, the critical strains of approximately 20\% indicate that the hydrogenated irida-graphene lattice combines a reduced stiffness with a considerable tolerance to tensile deformation, which suggests that irida-graphane could withstand the deformations involved in flexible device architectures without loss of structural integrity.

\begin{figure}[h!]
		\centering
		\includegraphics[width=\linewidth]{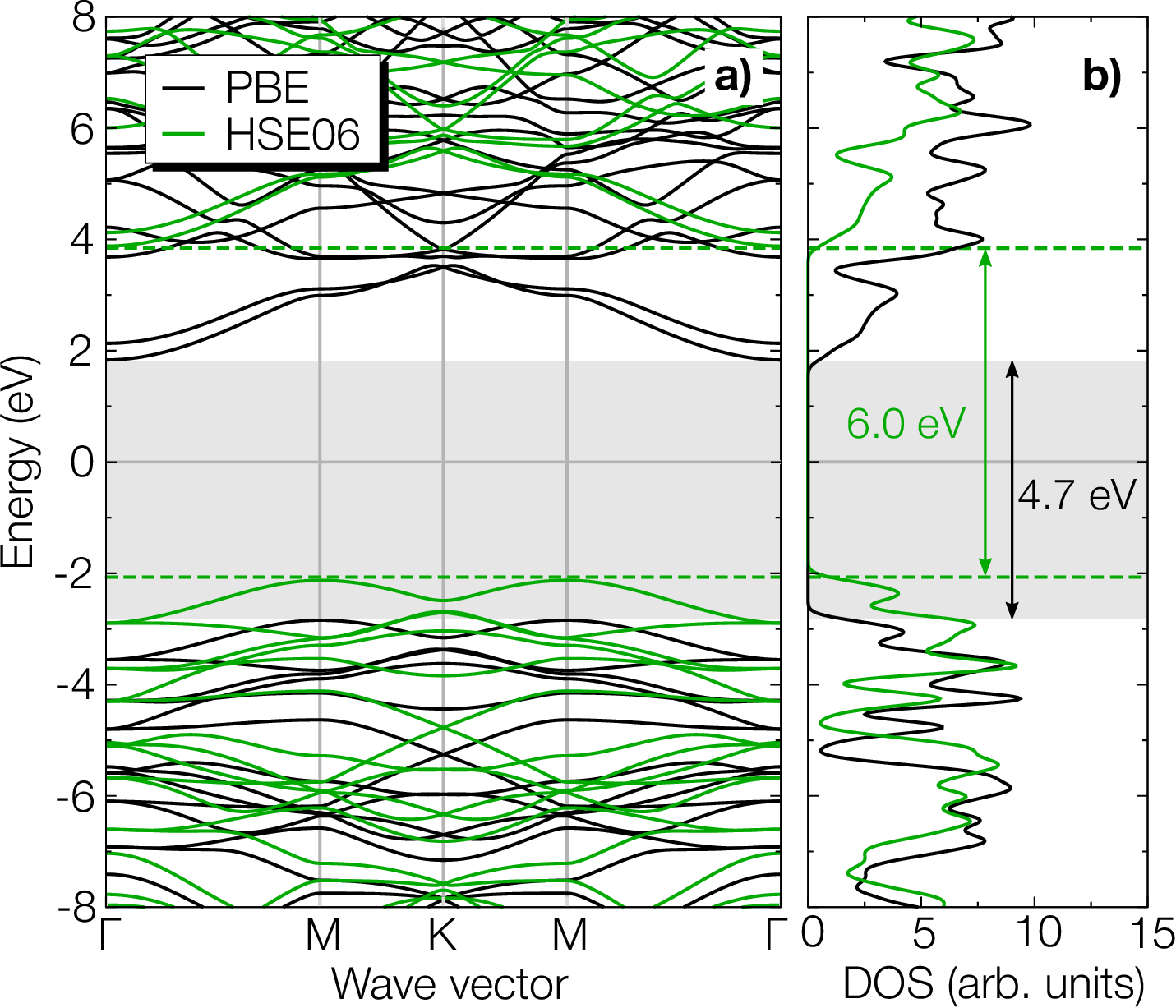}
		\caption{(a) Electronic band structure and (b) density of states of irida-graphane calculated at the PBE (black) and HSE06 (green) levels. The Fermi level is set to zero.}
		\label{fig:bands}
	\end{figure}

	The same saturation of the carbon framework that softens the lattice has an even more pronounced effect on its electronic structure. In graphene, hydrogenation removes the delocalized $\pi$ states and converts a semimetal into a wide-gap insulator \cite{PLACEHOLDER_graphane_prediction_1, PLACEHOLDER_graphane_experiment_1, PLACEHOLDER_graphane_gap_1}, and an analogous transformation is expected for the metallic irida-graphene \cite{PLACEHOLDER_irida_graphene_original_1}. The electronic band structure and the density of states (DOS) of irida-graphane calculated at the PBE and HSE06 levels are presented in Figure~\ref{fig:bands}. We observe that irida-graphane is a semiconductor with an indirect band gap, with the valence band maximum (VBM) located at the M point and the conduction band minimum (CBM) located at the $\Gamma$ point. The band gap is 4.68~eV at the PBE level and 6.0~eV with the HSE06 functional, and the smallest direct gap, located at the $\Gamma$ point, is 5.39~eV at the PBE level.

	This wide band gap is a direct consequence of the complete \textit{sp}$^{3}$ hybridization of the carbon framework. In pristine irida-graphene, the partially filled $\pi$ bands derived from the \textit{p}$_z$ orbitals give rise to its metallic character \cite{PLACEHOLDER_irida_graphene_original_1}. Upon hydrogenation, each \textit{p}$_z$ orbital forms a $\sigma$ bond with a hydrogen atom, and the band edges become determined by the bonding and antibonding $\sigma$ states, which are separated by a much larger energy. The band gaps of irida-graphane are larger than those reported for chair graphane, namely approximately 3.5~eV at the PBE level \cite{PLACEHOLDER_graphane_prediction_1, PLACEHOLDER_graphane_gap_1} and 4.5--4.6~eV with the HSE06 functional \cite{PLACEHOLDER_graphane_hse_gap_1}, which may be related to the longer C-C bonds and to the reduced orbital overlap in the strained ring network. Regarding the band dispersion, the conduction band is strongly dispersive around the $\Gamma$ point, with an energy increase of approximately 1.7~eV between $\Gamma$ and K, whereas the topmost valence band varies by approximately 0.7~eV along the path, which is reflected in the steep increase of the DOS at the valence band edge. The HSE06 functional increases the band gap by approximately 1.3~eV, as expected from the partial incorporation of screened exact exchange, which reduces the band gap underestimation of semilocal functionals \cite{PLACEHOLDER_pbe_gap_underestimation_2, PLACEHOLDER_hse06_3}, while preserving the overall dispersion of the bands. A band gap of this magnitude places irida-graphane among the wide-gap two-dimensional materials and is relevant for dielectric and ultraviolet optoelectronic applications.

\begin{figure}[h!]
		\centering
		\includegraphics[width=0.95\linewidth]{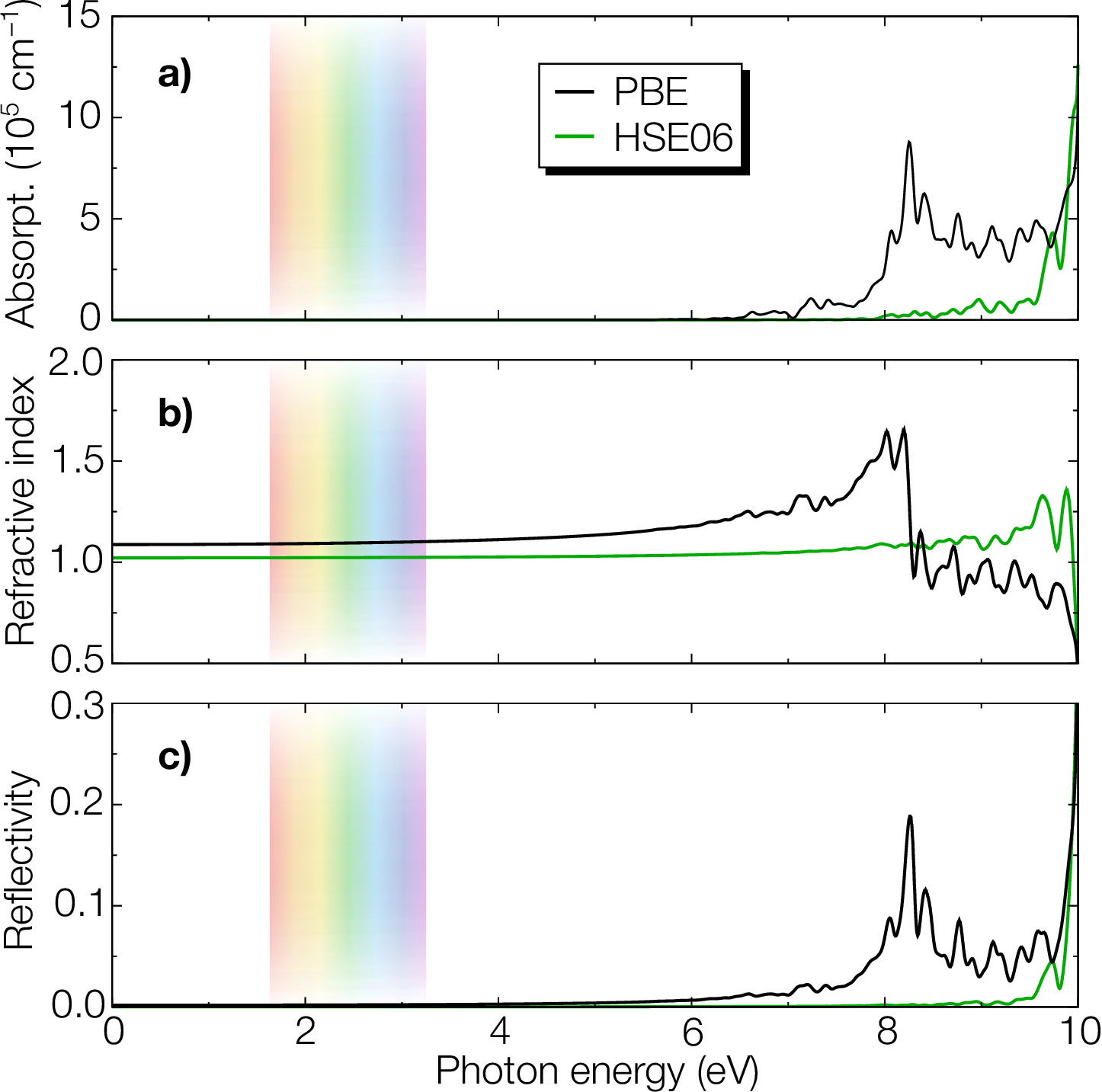}
		\caption{(a) Absorption coefficient, (b) refractive index, and (c) reflectivity of irida-graphane at the PBE (black) and HSE06 (green) levels. The shaded region indicates the visible range.}
		\label{fig:optic}
	\end{figure}

	Such a wide gap also implies that the optical activity of irida-graphane is confined to the ultraviolet region. To quantify this behavior, we have computed the dielectric function of the monolayer. Its imaginary part, which accounts for the interband optical transitions between the valence band (VB) and the conduction band (CB), is given by Fermi's golden rule \cite{tignon1995}
	\begin{equation}
		\label{eq:epsilon_2}
		\epsilon_2(\omega)=\frac{4\pi^2}{\Omega\omega^2} \sum_{\substack{i\in \text{VB}\\ j\in \text{CB}}} \sum_{k}W_k \left | \rho_{ij} \right |^2 \delta (\epsilon_{kj}-\epsilon_{ki}-\hbar\omega) \ \ ,
	\end{equation}
	\noindent where $\omega$ represents the photon frequency, $\left | \rho_{ij} \right |$ is the dipole transition matrix element, $W_k$ is the weight of the respective $k$-point in reciprocal space, and $\Omega$ is the system volume, while its real part follows from the Kramers-Kronig relation \cite{kramers1927, kronig1926}
	\begin{equation}
		\epsilon_1(\omega)=1+\frac{1}{\pi}P\int_{0}^{\infty}d\omega'\frac{\omega'\epsilon_2(\omega')}{\omega'^{2}-\omega^{2}} \ \ ,
	\end{equation}
	\noindent where $P$ denotes the principal value of the integral over $\omega'$. From $\epsilon_1$ and $\epsilon_2$, we have obtained the absorption coefficient ($\alpha$), reflectivity ($R$), and refractive index ($\eta$)
	\begin{equation}
		\alpha(\omega)=\sqrt{2}\omega\left [  (\epsilon_1^{2}(\omega)+\epsilon_2^{2}(\omega))^{1/2}-\epsilon_1(\omega)\right ]^{1/2} ,
	\end{equation}
	\begin{equation}
		R(\omega)=\left [ \frac{(\epsilon_1(\omega)+i\epsilon_2(\omega))^{1/2}-1}{(\epsilon_1(\omega)+i\epsilon_2(\omega))^{1/2}+1} \right ]^2 ,
	\end{equation}
	and
	\begin{equation}
		\quad \eta(\omega)=\frac{1}{\sqrt{2}}\left [ (\epsilon_1^{2}(\omega)+\epsilon_2^{2}(\omega))^{1/2}+\epsilon_1(\omega) \right ]^{1/2} ,
	\end{equation}
	\noindent which are presented in Figure~\ref{fig:optic} for in-plane polarized light at the PBE and HSE06 levels, where the shaded region indicates the visible range.

	As anticipated from the band gap, irida-graphane does not absorb in the infrared and visible ranges, as shown in Figure~\ref{fig:optic}(a). At the PBE level, the absorption begins at approximately 5.6~eV, which is close to the direct gap at the $\Gamma$ point and above the indirect gap, since the indirect transitions do not contribute to Eq.~\ref{eq:epsilon_2}. The absorption coefficient exceeds 10$^{5}$~cm$^{-1}$ at 7.2~eV, exhibits a pronounced peak of 8.8~$\times$~10$^{5}$~cm$^{-1}$ at 8.25~eV, and rises again to 1.1~$\times$~10$^{6}$~cm$^{-1}$ at the upper limit of the computed spectrum. With the HSE06 functional, the absorption onset shifts to approximately 6.7~eV, the absorption coefficient exceeds 10$^{5}$~cm$^{-1}$ only at 9.0~eV, and the strongest absorption occurs at the upper limit of the computed spectrum, consistent with the blue shift of the band edges discussed above.

	In the visible range, the refractive index shown in Figure~\ref{fig:optic}(b) remains nearly constant, between 1.09 and 1.10 at the PBE level, with static values of 1.09 (PBE) and 1.02 (HSE06). At the PBE level, it increases in the ultraviolet up to a maximum of 1.65 at 8.20~eV, just below the absorption peak at 8.25~eV, and then decreases, which is the typical dispersion behavior near a strong absorption band. Likewise, the reflectivity shown in Figure~\ref{fig:optic}(c) remains below 0.3\% in the visible range at both levels of theory and reaches 19\% at 8.26~eV at the PBE level, following the absorption peak, before rising steeply toward the upper limit of the spectrum. The combination of negligible absorption and reflectivity in the visible range with strong absorption in the deep ultraviolet suggests that irida-graphane may be suitable for transparent coatings and ultraviolet-filtering applications.

	\section{Conclusions}

	In summary, we have performed a first-principles investigation of the structural, mechanical, electronic, and optical properties of irida-graphane, the fully hydrogenated derivative of irida-graphene. Full hydrogenation converts the carbon atoms from \(sp^2\) to \(sp^3\) hybridization and produces a buckled framework in which the hydrogen atoms attached to the three- and six-membered rings lie on opposite sides of the monolayer. The three-membered rings retain saturated, cyclopropane-like configurations, while the buckling of the carbon framework reduces the in-plane projections of the C–C bonds, leading to lattice contraction despite their elongation. The calculated hydrogenation energy indicates an exothermic process relative to pristine irida-graphene and molecular hydrogen. Furthermore, the phonon dispersion exhibits no imaginary frequencies along the investigated high-symmetry path, and \textit{ab initio} molecular dynamics simulations show that the structure retains its integrity over 10~ps at temperatures up to 2000~K under the simulated conditions.

    Our calculations also reveal a wide indirect band gap, transforming metallic irida-graphene into a semiconductor with band gaps of 4.68~eV at the PBE level and 6.0~eV with the HSE06 functional. These values exceed those reported for chair graphane. The calculated mechanical response indicates that irida-graphane can sustain substantial tensile deformation, although its mechanical properties warrant further comparative analysis. The optical spectra predict negligible absorption and reflectivity in the visible range, together with strong absorption in the deep ultraviolet. Overall, these findings establish hydrogenation as a promising strategy for tailoring the structural and electronic properties of irida-graphene and identify irida-graphane as a candidate material for ultraviolet optoelectronics and transparent coatings. They also motivate further investigations into its experimental synthesis and the controlled hydrogenation of this carbon allotrope.

	\section*{Author Contributions}
	\noindent
	D.S.G.: Data Curation, Investigation, Software, Writing - Original Draft. I.M.F.: Investigation, Formal Analysis, Visualization, Writing - Original Draft. M.L.P.J.: Conceptualization, Methodology, Formal Analysis, Validation, Resources, Supervision, Project Administration, Funding Acquisition, Writing - Review \& Editing. All authors reviewed the manuscript.

	\section*{Declaration of Competing Interests}
	The authors declare that they have no known competing financial interests or personal relationships that could have appeared to influence the work reported in this paper.

	\section*{Data Availability}
	The data that support the findings of this study are available from the corresponding author upon reasonable request.

	\section*{Acknowledgments}
	D.S.G. acknowledges financial support from the Coordena\c{c}\~{a}o de Aperfei\c{c}oamento de Pessoal de N\'{i}vel Superior (CAPES). M.L.P.J. acknowledges financial support from FAPDF (grant 00193-00001807/2023-16), CNPq (grants 444921/2024-9 and 308222/2025-3), and CAPES (grant 88887.005164/2024-00).

	\bibliography{bibliography-new}

\end{document}